\documentclass[12pt]{iopart}
\usepackage{cite}

\usepackage{iopams}

\usepackage{graphicx}

\DeclareSymbolFont{AMSb}{U}{msb}{m}{n}
\DeclareSymbolFontAlphabet{\Bbb}{AMSb}

\newcommand{\B}{\mathcal{B}}
\newcommand{\N}{\Bbb{N}}

\newcommand{\R}{\Bbb{R}}
\newcommand{\C}{\Bbb{C}}

\newcommand{\si}{\sigma_I}
\newcommand{\sii}{\sigma_{II}}
\newcommand{\li}{\lambda_I}
\newcommand{\lti}{\tilde{\lambda}_I}
\newcommand{\lii}{\lambda_{II}}
\newcommand{\ltii}{\tilde{\lambda}_{II}}
\newcommand{\ei}{\epsilon_I}
\newcommand{\eii}{\epsilon_{II}}
\newcommand{\rhoi}{\rho_{\phi_I}}
\newcommand{\rhoii}{\rho_{\phi_{II}}}
\newcommand{\oi}{\omega_{\phi_I}}
\newcommand{\oii}{\omega_{\phi_{II}}}
\newcommand{\ot}{\tilde{\omega}}
\newcommand{\ii}{\mathrm{i}}

\newcommand{\ie}{i.e., }
\newcommand{\eg}{e.g., }
\newcommand{\eqref}[1]{(\ref{#1})}

\begin{document}
20 November 2003
{\hspace*{\fill} Preprint-KUL-TF-2003/32}

\title{A microscopic model for Josephson currents}
\author{J.Lauwers, A.Verbeure}

\address{Instituut voor Theoretische Fysica,  
Katholieke Universiteit Leuven,   
Celestijnenlaan 200D,   
B-3001 Leuven, Belgium}
\ead{\mailto{joris.lauwers@fys.kuleuven.ac.be}, \mailto{andre.verbeure@fys.kuleuven.ac.be}}

\begin{abstract}
A microscopic model of a Josephson junction between two 
superconducting plates is proposed and analysed.  For this model,
the nonequilibrium steady state of the total system is explicitly 
constructed and its properties are analysed.
In particular, the Josephson current is rigorously computed as a function of 
the phase difference of the two plates and the typical properties of the Josephson 
current are recovered.
\end{abstract}


\section{Introduction}
In 1962, Josephson \cite{josephson:1962} predicted a novel phenomenon in
superconductivity, namely when two different superconductors were brought into 
close contact. Based on elementary quantum mechanics, he predicted the 
existence of a supercurrent with a peculiar current-voltage dependence. 
Namely, he argued that there would emerge a current of Cooper pairs which is 
proportional to the sine of the phase difference of the order parameters of both
superconductors.
The success of this prediction was immediate when indeed this phenomenon was 
experimentally observed already one year later \cite{anderson:1963}. 
It counts as one of the greatest 
successes of quantum mechanics in physics and you will find a chapter on the 
Josephson effects in almost every textbook on superconductivity.  
During the following decades,
the increase of knowledge on this subject in theoretical solid state physics 
has been tremendous and applications of Josephson junctions 
in electronic devices have been developed \cite{likharev:1986}.
Progress in conceiving a microscopic theory for the Josephson effect 
in rigorous quantum statistical mechanics was made when Sewell obtained the 
Josephson and Meissner effects in an model independent approach from the 
assumption of offdiagonal long range order and local gauge covariance 
\cite{sewell:1997}.

Here, we are concerned with a substantive and rigorous derivation of 
the Josephson current in a concrete microscopic quantum statistical model. 
The model consists of two twodimensional superconducting plates which have a 
common onedimensional contact surface through which Cooper pairs can tunnel, 
and as such it induces a current. We use recently developed ideas of 
nonequilibrium statistical mechanics to conceive a reasonable framework for 
this model in which the Josephson current can be computed explicitly.  

The model at hand is based on the mean field version, sometimes called the
Anderson version, of the strong coupling 
BCS model. This model may lack many physical features, it is nevertheless an 
interesting model as it allows rigorous and explicit computations 
leading to nontrivial results.

Due to the phase difference between the superconductors, the total system is 
not in an equilibrium state and we have to find a reasonable description of the
nonequilibrium steady state of the system. 
In spite of the great success of statistical mechanics for systems 
in equilibrium, still not very much is understood in the nonequilibrium 
situation, 
especially for quantum systems. One of the most important points in this note 
is thus the construction of a good nonequilibrium steady state for this model
(Section~\ref{Sness}).
This nonequilibrium steady state is constructed by selecting the state which 
fulfils a list of natural conditions in this setup. This construction
is then compared with other treatments found in the literature 
\cite{ruelle:2000,ruelle:2001}. 

After developing our framework for this model, we calculate 
the currents in the system (Section~\ref{Scur}). 
First, we find that there is a Cooper pair current with the typical phase 
dependence of the traditional Josephson currents (see Figure~\ref{fig3}). 
Secondly, this model can be compared with other 
nonequilibrium quantum statistical models for entropy production  
\cite{ruelle:2001,jaksic:2002,aschbacher:2003,matsui:2003} 
and we also compute the heat fluxes between the two superconductors. 
In spite of the fact that there are nontrivial particle currents in this system, 
we get that there are no heat fluxes or entropy production here.

\section{The microscopic model}\label{sectBCS}

\begin{figure}[h]
\begin{center}
\includegraphics{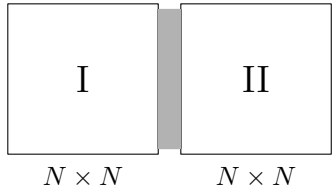}
\caption{Two superconductors with a contact surface}
\label{fig1}
\end{center}
\end{figure}

The model we propose consists of two superconducting plates, $I$ and $II$. 
At one side, the distance between these superconductors is made very small, 
such that Cooper pairs can tunnel through this barrier, see
Figure~\ref{fig1}, \ie the superconducting plates have a common contact surface.  

We conceive a microscopic model for this setup and describe the
superconductors by square lattices, \ie $\N^2 \oplus \N^2$, where the distance 
between the superconductors is of the order of the lattice site distance.
The  interactions in this system are given by local Hamiltonians $H_N$, \ie for
a finite lattice with $N^2 + N^2$ points,
\begin{equation}\label{HN}
H_N = H_{I,N} + H_{II,N} + V_N.
\end{equation}
Here, $H_{I,N}$ and $H_{II,N}$ are the Hamiltonians of the two
superconductors, labelled by $I$ and $II$, and $V_N$ models the interaction of 
the junction between the two superconductors. It is important to realise that
the individual superconducting plates are twodimensional and that the contact is
onedimensional, but nevertheless infinite in the thermodynamic limit, \ie the
limit $N \to \infty$.

The superconductors are modelled by the strong-coupling BCS model, using the 
quasi-spin formalism \cite{thirring:1967,thirring:1968}, \ie the 
interactions in the superconductors are given by the Hamiltonians $H_{I,N}$ and
$H_{II,N}$ for finite $N\in\N$,
\begin{equation}\label{BCSHAM}\fl
H_{i,N} = \sum_{k,l=0}^{N-1}\ei \sigma_i^z(k,l) - 
\frac{1}{N^2}\sum_{k,l,m,n = 0}^{N-1}\sigma_i^+(k,l)
\sigma_i^-(m_1,n_1), \quad \epsilon_i > 0,\ i = I,II.
\end{equation}
These Hamiltonians $H^N_i$ act on the Hilbert space 
$\bigotimes_{j= 0}^{(N-1)^2} \C_j^2$; $\sigma_i^+$, $\sigma_i^-$ and 
$\sigma_i^z = \sigma_i^+\sigma_i^- - \sigma_i^-\sigma_i^+$ are copies 
of the well-known Pauli matrices, generators of the algebra of $2 \times 2$
complex matrices $M_2$.
 The operators $\si^+(k,l)$ and $\si^-(k,l)$
represent respectively the creation and annihilation operators for a Cooper
pair in superconductor $I$ with indices $(k,l)$; the $\epsilon_i$ are the
kinetic energies of the Cooper pairs.   

The junction between the two superconductors $I$ and $II$ is modelled by 
\begin{equation}\label{coupl}\fl
V_N = - \frac{\gamma}{N} \sum_{k_1,k_2 = 0}^{N-1} \si^+(k_1,0)\sii^-(k_2,0)
+\si^-(k_1,0)\sii^+(k_2,0), \quad \gamma \in \R^+. 
\end{equation}
It describes the rate by which Cooper pairs tunnel through
the barrier (Fig.~\ref{fig1}). A Cooper pair with indices $(k_1,0)$ in the 
first superconductor can tunnel through the junction and become a Cooper pair 
with indices $(k_2,0)$ in the second superconductor. 
The coupling constant $\gamma$ governs 
the rate at which these processes  occur.  
Note that only Cooper pairs which are on the
contact surface, \ie pairs which have a second index equal to zero, can tunnel
through the junction.   The BCS model \eqref{BCSHAM} we use here is
permutation invariant with respect to the lattice indices. The underlying
geometry of the lattices plays no further role. The important point is that 
only $N$ out of $N^2$ sites of one superconductor, see equation \eqref{coupl}, 
are linked with the other superconductor. The high degree of permutation 
symmetry in the system makes the model exactly solvable in the thermodynamic 
limit, $N\to\infty$.

\subsection{Equilibrium properties of the noninteracting
superconductors\label{eq}}
Before studying the coupling between the two superconductors \eqref{coupl}, we 
briefly describe the equilibrium properties and the phase transition in the 
noninteracting superconductors, \ie the situation $\gamma=0$. We treat 
the first superconductor explicitly, the second is analogous.

Exploiting the permutation invariance of
the BCS Hamiltonians \eqref{BCSHAM} with respect to the lattice indices, the properties of the equilibrium
states in the thermodynamic limit, $N\to\infty$, of this model can elegantly
be derived using \cite{fannes:1980}.
The extremal $\beta_I$-KMS or equilibrium states at inverse temperature 
$\beta_I$ in the thermodynamic limit are given by the product states $\oi$, 
where the expectation value of 
all $X = X_0 \otimes X_1 \otimes X_2 \otimes \cdots$ in the 
infinite tensor product algebra of local observables, $\mathcal{B}_I = 
\overline{\bigcup_{B \subset \N^2} \bigotimes_{i\in B} M_{2,i}}$, is given by:
\begin{equation}\label{prod}
\omega_{\phi_I}(X) =\prod_{i\in\N^2} \tr \rho_{\phi_I,i} X_i.
\end{equation}
Here, $\rho_{\phi_I,i} $ is the $i$-th copy of the $2 \times 2$ density matrix
$\rho_{\phi_I} \in M_2$, a solution of the selfconsistency or gap equation:
\begin{equation}\label{gap1}
\rho_{\phi_I} = \frac{\exp[-\beta_I h_{\phi_I}]}{\tr \exp[-\beta_I
     h_{\phi_I}]},
\end{equation}
where $h_{\phi_I}$ is a one-site effective or selfconsistent 
Hamiltonian,
   \begin{equation}\label{heff}
     h_{\phi_I} = \ei\si^z - \lambda_I e^{i\phi_I}\si^- - \lambda_I
     e^{-i\phi_I}\si^+, 
   \end{equation}
with an order parameter $\lambda_I$, satisfying the selfconsistency equation:
 \begin{equation}\label{lI} 
     \lambda_I  =  \left| \tr \rhoi\si^+\right |= \left| \oi
     (\si^+)\right |;  
 \end{equation}
derived from equations \eqref{prod} and \eqref{gap1},
and with a phase $\phi_I$:
 \begin{equation}\label{fI}
    \phi_I = \arg \tr \rhoi\si^+ = \arg \oi(\si^+),
    \qquad   \phi_I \in [0, 2\pi].
 \end{equation}
By explicit calculation, the gap equation \eqref{lI} can be rewritten in 
the form
      \begin{equation}\label{gap2}
        \lambda_I (1 - \frac{1}{2k_I}\tanh[\beta_I k_I]) = 0,  
      \end{equation}
where $k_I = \sqrt{\ei^2 + \lambda^2_I}$ and $\{-k_I,k_I\}$ is the spectrum of 
the effective Hamiltonians $h_{\phi_I}$ (\ref{heff}), and is 
independent of the phase $\phi_I$.
It can readily be seen that the equation (\ref{gap2}) has always a solution
$\lambda_I =0$. It yields the normal phase state. 
Solutions $\lambda_I \ne 0$
exist if the following conditions are satisfied:
\begin{equation}\label{crit}
\left \{ 
   \begin{array}{lcl} 
     \ei & < & 1/2, \\
     \beta_I & > & \beta_c = 
     \frac{1}{2\ei}\log\left(\frac{1+2\ei}{1-2\ei}\right).
   \end{array} 
   \right .
\end{equation} 
These solutions $\lambda_I \ne 0$, $\phi_I \in [0, 2\pi]$ are understood to describe
the superconducting phase states. 
The order parameter $\li \ne 0$ \eqref{lI} depends on the inverse temperature
$ \beta_I$ which has to be larger than the critical inverse 
temperature for the BCS model $\beta_c$, cf.~\eqref{crit}.
The phase of the superconductor $\phi_I$ \eqref{fI} can be fixed freely. 
This leads to an infinite degeneracy of the states in the 
superconductive regime and is due to spontaneous symmetry breaking 
\cite{goderis:1991,michoel:2001,lauwers:2001b}.
The second superconductor $II$ has analogous normal and superconducting
equilibrium states.

\subsection{The nonequilibrium steady state (NESS)\label{Sness}}

Suppose that initially the two superconductors are both in a
superconducting pure phase equilibrium state at inverse
temperature $\beta_I$ and phase $\phi_I$ for the first BCS superconductor 
and inverse temperature $\beta_{II}$ and phase $\phi_{II}$ for 
BCS superconductor $II$. I.e., the state of this system is the product of two 
pure phase states \eqref{prod},  $\oi \otimes \oii$ on the total algebra 
of local observables $\B = \overline{\bigcup_{B \subset \N^2\oplus \N^2}
\bigotimes_{i\in B}M_{2,i}}$.

If the interactions between the two superconductors are switched on, the system
should ultimately evolve to a nonequilibrium steady state $\ot$ for the 
dynamics of the total interacting system \eqref{HN}, \ie 
\begin{equation}\label{ss0}
\ot = \lim_{t\to\infty} \oi \otimes \oii \circ \alpha_t
\end{equation}
where $\{\alpha_t\}$ is the dynamical semigroup of the total interacting infinite
system, \ie
\begin{equation}\label{dyn}
\alpha_t(.) = \lim_{N\to \infty} \e^{\ii t H_N} . \ \e^{- \ii t H_N}.
\end{equation}
However, it is not clear that the limit \eqref{ss0} exists
for our model \eqref{HN}. Computer simulations
suggest that the system continues to evolve periodically in time. Furthermore, 
it has to be specified in which sense the limiting dynamics $\alpha_t$ 
\eqref{dyn} can be defined.

To deal with these problems, Ruelle proposed the following construction for a
natural nonequilibrium steady state \cite{ruelle:2000,ruelle:2001}:
\begin{equation}\label{ness}
\tilde{\omega} = \lim_{T\to\infty} \frac{1}{T} \int_0^T \! \mathrm{d}t\
\omega_0 \circ \alpha_t,
\end{equation} 
where $\omega_0$ is the initial equilibrium state.
Ruelle considered the case of a finite system coupled to a number of infinite
reservoirs, and the conditions on the interactions between the small system and
the reservoirs where such that the limiting dynamics \eqref{dyn} exists 
in norm on the $C^*$-algebra, \ie independent of the state of the system.
 Clearly, the state $\tilde{\omega}$ \eqref{ness} is an invariant state for the dynamics of the
total interacting system \eqref{dyn} and it is called a natural nonequilibrium
steady state. In recent years, exact results concerning
entropy production in quantum statistical models were derived using 
Ruelle's framework \eqref{ness}, see \eg 
\cite{ruelle:2001,jaksic:2002,aschbacher:2003,matsui:2003}.

In our model \eqref{HN}, the interactions are not local and the existence and
properties of the dynamics \eqref{dyn} depend on the state of the system. 
Hence, another approach has to be followed to obtain a nonequilibrium 
steady state. For this model, we propose to construct a steady state by 
selecting a good state which satisfies all the essential properties 
a NESS should have. Which are those properties?

\begin{figure}[h]
\begin{center}
\includegraphics{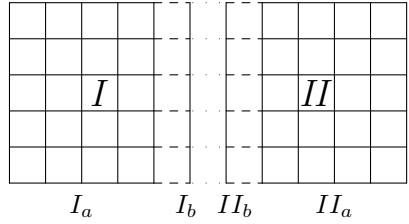}
\caption{Devision of the system into four permutation invariant subsystems}
\label{fig2}
\end{center}
\end{figure}

First, due to the high degree in permutation symmetry in the model \eqref{HN}, it is
natural to look for steady states in the class of product states, \ie
\begin{equation}\label{NESSprod}\fl
\ot(X) = \!\!\prod_{i \in \N^2\oplus \N^2}\!\! \tr \tilde{\rho}_i X_i, \quad X =
X_{k} \otimes X_{l}  \otimes X_{m} \otimes \cdots \in \B; X_j \in M_2.
\end{equation}
Furthermore, we expect the state $\ot$ to have the same permutation symmetries 
as the total Hamiltonian \eqref{HN}. Analysing this, we see that the total system
should be divided into four parts (see Figure~\ref{fig2}) 
which are permutation invariant subspaces. 
The first part is the bulk term of the
first superconductor $(I_a)$, the second piece is the contact or surface part of
the first superconductor $(I_b)$, the third part is the contact part of the
second superconductor $(II_b)$ and the fourth part is the bulk part of
the second superconductor $(II_a)$. 
The size of the bulk subsystems $(I_a,II_a)$ is of the order of $O(N(N-1))$ 
and the subsystems on the contact surface $(I_b,II_b)$ are of the order of 
$O(N)$. Hence, the states $\ot$ are  products 
of four symmetric product states on these different regions:
\begin{equation*}
\ot = \ot_{I_a}\otimes\ot_{I_b}\otimes\ot_{II_b}\otimes\ot_{II_a}.
\end{equation*}

Secondly,
requiring that $\ot$ is a steady state is mathematically translated into the
following expression:
\begin{equation*}
\lim_{N\to\infty}\ot\big(\big[H_N,X\big]\big) = 0,\quad \forall X \in \B.
\end{equation*}
Due to the product structure of the steady state \eqref{NESSprod}, the limit
$N\to\infty$ exists; 
it yields an equivalent Hamiltonian of the type $\tilde{H}_N = \sum_i
\tilde{h}_i$, and the condition for the time invariance becomes now:
\begin{equation}\label{NESSsteady}
\ot([\tilde{h}_i,X_i]) = 0,\quad \forall X_i \in M_{2,i},
\end{equation}
where $\tilde{h}_{i}\in M_{2,i}$ is the site 
dependent effective one-site Hamiltonian for site $i \in 
\N^2 \oplus \N^2$. It follows directly from the permutation 
symmetries of $H_N$ \eqref{HN} and the product structure of 
the state $\ot$ \eqref{NESSprod} that these effective Hamiltonians have 
the following structure:
\begin{equation}\fl\label{htilde}
\tilde{h}_i = \left \{
 \begin{array}{ll}
 \ei\si^z(i) -\ot(\sigma^+_{I_a})\sigma_{I_a}^-(i) -
 \ot(\sigma^-_{I_a})\sigma_{I_a}^+(i), 
 & i \in I_a;\\[0.3cm]
 \ei\sigma_{I_b}^z(i) -\ot(\sigma^+_{I_a})\sigma_{I_b}^-(i) - \ot(\sigma^-_{I_a})\sigma_{I_b}^+(i)
 & \\ \qquad\qquad\qquad
 -\gamma \Big( \ot(\sigma^+_{II_b})\sigma^-_{I_b}(i) + \ot(\sigma^-_{II_b})\sigma^+_{I_b}(i) 
 \Big),& i \in I_b;\\[0.3cm]
 \eii\sigma_{II_b}^z(i) -\ot(\sigma^+_{II_a})\sigma_{II_b}^-(i) - 
 \ot(\sigma^-_{II_a})\sigma_{II_b}^+(i)
 & \\ \qquad\qquad\qquad
 -\gamma \Big( \ot(\sigma^+_{I_b})\sigma^-_{II_b}(i) + \ot(\sigma^-_{I_b})
 \sigma^+_{II_b}(i)\Big),& i \in II_b;\\[0.3cm]
 \eii\sigma_{II_b}^z(i) -\ot(\sigma^+_{II_a})\sigma_{II_b}^-(i) - 
 \ot(\sigma^-_{II_a})\sigma_{II_b}^+(i),
  &i\in II_a,
 \end{array} \right .
\end{equation}
where $\sigma^\#_{t}$ is the $t$-th copy of the Pauli matrix $\sigma^\#$ in the 
site $t \in \{I_a,I_b,II_a,II_b\}$. The expectation
values $\ot(\sigma^\pm_{t})$ appearing in these expressions 
are the order parameters for this effective Hamiltonian. They are 
determined in a selfconsistent way, see below, by the effective Hamiltonian $\tilde{h}$ 
\eqref{htilde} and the invariant state $\ot$ \eqref{NESSsteady}. 

A third requirement is that a NESS $\ot$ should have the same properties as the
initial equilibrium state $\oi \otimes \oii$ if we look at the system
far enough away from the contact surface, \ie 
the expectation values for local observables 
$X \in \B$ with a support in $I_a$ or $II_a$ satisfy:
\begin{equation}\label{NESSnat}
\begin{array}{l}
\ot(X) = \oi(X), \quad \mathit{for}\ X\mathit{\ with\ support\ in\ }I_a;\\[0.1cm]
\ot(X) = \oii(X), \quad \mathit{for}\ X\mathit{\ with\ support\ in\ }II_a.
\end{array}
\end{equation}
In this sense, the steady state $\ot$ has a memory of the initial equilibrium 
state $\oi \otimes \oii$, a property which seems indeed a natural ingredient for
a NESS.

The NESS can now be constructed explicitly as follows.
The requirement that $\ot$ is an invariant product state \eqref{NESSsteady} 
for the dynamics induced by $\tilde{h}$ \eqref{htilde} leads to the 
following choice for the one-site density matrices  $\tilde{\rho}_i$ of $\ot$ 
\eqref{NESSprod}:
\begin{equation*}
\tilde{\rho}_i = c_+|\tilde{+}\rangle \langle \tilde{+}|_i  
+ c_-|\tilde{-}\rangle \langle \tilde{-}|_i, 
\quad \forall i \in 
\N^2  \oplus  \N^2,
\end{equation*}
where $|\tilde{+}\rangle_i$ and $|\tilde{-}\rangle_i$ are 
the eigenvectors of $\tilde{h}_i$, and the constants $c_-$ and 
$c_+$ are two positive constants adding up to one. 
A natural way to implement condition \eqref{NESSnat} is to 
to define the constants $c_-$ and $c_+$ by means of the expectation values 
of the one-site density matrices of the initial equilibrium state 
$\oi \otimes \oii$ on the eigenstates of $\tilde{h}_i$, \ie we chose
\begin{equation*}
c_+= \langle \tilde{+}|\rho_{0,i}|\tilde{+}\rangle_i, \quad \mathrm{and}, \quad
c_-= \langle \tilde{-}|\rho_{0,i}|\tilde{-}\rangle_i.
\end{equation*} 
In other words, for every site $i$ in both superconductors, we project the 
local one-site density matrices $\rho_{0,i}$ of the initial equilibrium 
state $\oi \otimes \oii$ onto the subalgebra in $M_2$ which is 
invariant under the dynamics induced by $\tilde{h}$ \eqref{htilde}.
Hence, the one-site density matrices $\tilde{\rho}_i$ 
of the NESS are given by
\begin{equation}\label{ness2}
\tilde{\rho}_i = |\tilde{+}\rangle \langle \tilde{+}| \langle
\tilde{+},\rho_{0}\tilde{+}\rangle_i + |\tilde{-}\rangle \langle \tilde{-}| 
\langle \tilde{-},\rho_{0}\tilde{-}\rangle_i,\quad \forall i \in 
\N^2  \oplus  \N^2.
\end{equation}
Furthermore, in order to meet condition \eqref{NESSnat} and to 
avoid trivial nonsuperconducting solutions in \eqref{ness2}, we explicitly set 
the bulk order parameters and phases 
equal to the equilibrium order parameters and phases, \ie
\begin{equation}\fl \label{op1}
\ot(\sigma^+_{I_a}) = \li \e^{\ii \phi_I} = \oi(\si^+), \quad \mathrm{and}, \quad
\ot(\sigma^+_{II_a}) = \li \e^{\ii \phi_{II}}= \oii(\sii^+).
\end{equation}

It is easy to check that this choice \eqref{ness2}--\eqref{op1} yields a state $\ot$ 
fulfilling indeed all of the above requirements.
By construction, this state is again a product state \eqref{NESSprod}
and it is an invariant state under the limiting dynamics \eqref{NESSsteady}. 
In sites which do not belong to the contact surface, \ie $i \ne (k_1,0) \in I_b$ and $i \ne (k_2,0)
\in II_b$, the local density matrices of the initial equilibrium state and 
the constructed NESS coincide \eqref{NESSnat}.  
This can be seen as follows. In the bulk regions, \ie  for sites 
$i \in I_a \cup II_a$, the effective Hamiltonians $\tilde{h}_i$ \eqref{htilde}
coincide with the  equilibrium effective Hamiltonians $h_{\phi_I}$ or
$h_{\phi_{II}}$ \eqref{heff}. Hence in $I_a \cup II_a$, 
the spectral decompositions of $\rhoi$ or
$\rhoii$ coincide with those given in \eqref{ness2}.

Now, we have to compute the order parameters and phases in 
the effective Hamiltonians \eqref{htilde}. 
Compared to the equilibrium
situation, see section~\ref{eq}, 
where two order parameters \eqref{lI} and two phases \eqref{fI}
completely determine the equilibrium states, we have now
four different order parameters and four phases appearing, cf.~\eqref{htilde}, 
\ie apart from the parameters in Equation \eqref{op1}, we also have to 
compute
\begin{eqnarray*}
\tilde{\lambda}_I = |\tilde{\omega}(\sigma^+_{I_b})|,
&&\quad \mathrm{and}, \quad
\tilde{\phi_I} = \arg \tilde{\omega}(\sigma^+_{I_b}); 
\\
\tilde{\lambda}_{II} = |\tilde{\omega}(\sigma^+_{II_b})|,
&&\quad \mathrm{and}, \quad
\tilde{\phi_{II}} = \arg \tilde{\omega}(\sigma^+_{II_b}),
\end{eqnarray*}
Using \eqref{ness2} and \eqref{op1}, we find an explicit expression for the
selfconsistency equations in the steady state $\ot$ which determine these order
parameters and phases:  
\begin{eqnarray} \nonumber
 \lo{\tilde{\lambda}_I \e^{\ii\tilde{\phi}_I}  =}  \tilde{\omega}(\si^+(k_1,0)) 
\\ \lo= 
 (\lambda_I\e^{\ii\phi_I} + \gamma\tilde{\lambda}_{II}\e^{\ii\tilde{\phi}_{II}})
 \frac{\ei^2 + \lambda_I|\lambda_I \e^{\ii\phi_I} + \gamma \tilde{\lambda}_{II}
 \e^{\ii\tilde{\phi}_{II}}|\cos(\phi_I -\tilde{\phi}_I)}{\ei^2 + |\lambda_I
 \e^{\ii\phi_I} + \gamma\tilde{\lambda}_{II}\e^{\ii\tilde{\phi}_{II}}|^2} 
 \label{op2}
\\ \nonumber
 \lo{\ltii \e^{\ii\tilde{\phi}_{II}} =}  \tilde{\omega}(\sii^+(k_2,0)) 
\\ \lo=
 (\lii\e^{\ii\phi_{II}} + \gamma\tilde{\lambda}_{I}\e^{\ii\tilde{\phi}_{I}})
 \frac{\eii^2 + \lii|\lii \e^{\ii\phi_{II}} + \gamma \tilde{
 \lambda}_{I} \e^{\ii\tilde{\phi}_{I}}|\cos(\phi_{II} -\tilde{\phi}_{II})}
 {\eii^2 + |\lii\e^{\ii\phi_{II}} + \gamma\lti
 \e^{\ii\tilde{\phi}_{I}}|^2}
 \label{op3}
\end{eqnarray}
Together with the selfconsistency equations for $\lambda_I$ and $\lambda_{II}$
\eqref{gap2}, the above equations form a set of six coupled 
transcendental equations whose solutions completely determine the state 
$\tilde{\omega}$ \eqref{ness2}. 

The steady state $\tilde{\omega}$ \eqref{ness2} divides the system into four
parts. In both superconductors away from the contact surface, see 
Fig.~\ref{fig2}, the presence of the other superconductor is not felt. The 
system is unperturbed and behaves as in the initial equilibrium state.  
Nevertheless, the NESS \eqref{ness2} is not normal with respect to the 
equilibrium state $\oi \otimes \oii$, because the one-site density matrices 
$\tilde{\rho}_i$ on an infinite amount of sites on the contact surface 
(\ie the sites $i \in I_b \cup II_b$) differ from the one-site
density matrices of the equilibrium state \eqref{gap1}.  
On the contact surface, the system is effectively disturbed and there appear 
two intermediate layers with different properties compared with the 
initial equilibrium state. 

Finally, let us compare this construction of a NESS with the construction 
proposed by Ruelle \eqref{ness}.
The essential relation between our construction \eqref{ness2} 
and Ruelle's construction  \eqref{ness} is that in the latter one, 
it can be expected that the parts of the algebra which are not invariant under 
the dynamics \eqref{dyn} are averaged out and that the NESS is effectively a 
projection of the initial state on the subalgebra which is invariant under the 
dynamics, \ie it leads to the same state as in \eqref{ness2}.

This finishes the construction of our framework. Let us summarise the situation
at hand. We start with the microscopic model \eqref{HN} of two different
superconductors interacting via a particular Josephson junction and we
constructed a particular reasonable nonequilibrium steady state $\ot$ with site local
density matrices given by Equations \eqref{ness2}--\eqref{op3}.  

\section{Transport phenomena\label{Scur}}

In the framework described above, we can investigate the
nonequilibrium properties and the existence of currents which emerge by 
bringing two superconductors in contact with each other. In particular, we 
compute the expression for the Josephson particle current.

Suppose $Q_N$ is an extensive observable, \eg the total number of particles, 
the energy,\ldots The
transport related to this quantity $j(Q)$ in the steady state \eqref{ness2} 
is defined as,
\begin{equation}\label{curr}
j(Q) = \lim_{N\to\infty}\ii\frac{1}{N}\tilde{\omega}\left([H_N,Q_N]\right),
\end{equation}
where $H_N$ is the Hamiltonian of the total interacting system \eqref{HN}.
In this definition, the current is scaled with 
the size of the contact surface, \ie we divide by $N$. This
is the right scaling to obtain density quantities. Therefore, $j(Q)$ is the
current density.
Taking into account the product character of the steady state \eqref{ness2} 
and the fact that this state is invariant only under the limiting dynamics 
\eqref{NESSsteady}, we compute
\begin{eqnarray*}\fl
\ii\frac{1}{N} \tilde{\omega}([ H_N,Q_N]) = \ii\ \tilde{\omega}\Big ( \Big[ 
\sum_{k_1,l_1=0}^{N-1}\ei \si^z(k_1,l_1) 
+ \sum_{k_2,l_2=0}^{N-1} \eii \sii^z(k_2,l_2)
\\
- \frac{N(N-1)}{N^2} \sum_{k_1,l_1=0}^{N-1} \si^+(k_1,l_1) \li \e^{-\ii\phi_I} + 
\si^-(k_1,l_1)\li\e^{\ii\phi_I} \\
- \frac{1}{N}\sum_{k_1,l_1=0}^{N-1} \si^+(k_1,l_1) \lti \e^{-\ii\tilde{\phi}_I} + 
\si^-(k_1,l_1)\lti\e^{\ii\tilde{\phi}_I}
\\
-\gamma \sum_{k_1=0}^{N-1} \si^+(k_1,0) \ltii \e^{-\ii\tilde{\phi}_{II}} + 
\si^-(k_1,0)\ltii\e^{\ii\tilde{\phi}_{II}}
\\
- \frac{N(N-1)}{N^2} \sum_{k_2,l_2=0}^{N-1} \sii^+(k_2,l_2) \lii \e^{-\ii\phi_{II}} + 
\si^-(k_2,l_2)\lii\e^{\ii\phi_{II}}\\
- \frac{1}{N}\sum_{k_2,l_2=0}^{N-1} \sii^+(k_2,l_2) \ltii \e^{-\ii\tilde{\phi}_{II}} + 
\sii^-(k_2,l_2)\ltii\e^{\ii\tilde{\phi}_{II}}
\\
-\gamma \sum_{k_2=0}^{N-1} \sii^+(k_2,0) \lti \e^{-\ii\tilde{\phi}_{I}} + 
\sii^-(k_2,0)\lti\e^{\ii\tilde{\phi}_{I}},Q_N\Big]\Big ).
\end{eqnarray*}
This is simplified using that $\tilde{\omega}$ is invariant under the 
limiting dynamics, \ie $\tilde{\omega}([\sum_i\tilde{h}_i,Q_N])=0$
 \eqref{NESSsteady}. 
 We obtain
\begin{equation}\fl\label{transp}
 \ii\frac{1}{N} \tilde{\omega}([H_N,Q_N]) = 
 \ii\frac{1}{N^2} \tilde{\omega}\Big( \Big[ -\sum_{k_1,l_1=0}^{N-1} X_I(k_1,l_1)
 -\sum_{k_2,l_2=0}^{N-1} X_{II}(k_2,l_2) ,Q_N \Big] \Big),
\end{equation}
where
\begin{equation*}\fl
X_I(k_1,l_1) = \si^+(k_1,l_1)(\lti\e^{\ii\tilde{\phi}_{I}}-\li\e^{\ii\phi_{I}})
 + \si^-(k_1,l_1)(\lti\e^{-\ii\tilde{\phi}_{I}} - \li\e^{-\ii\phi_{I}}),
\end{equation*}
and
\begin{equation*}\fl
X_{II}(k_2,l_2) = \sii^+(k_2,l_2)(\ltii\e^{\ii\tilde{\phi}_{II}}-
\lii\e^{\ii\phi_{II}}) + \sii^-(k_2,l_2)(\ltii\e^{-\ii\tilde{\phi}_{II}} 
- \lii\e^{-\ii\phi_{II}}).
\end{equation*}
The order parameters and phases, $\li,\lti,\ldots$, appearing in these 
expressions are defined in Equations \eqref{gap2}, \eqref{op2} and \eqref{op3}.
Already from this expression \eqref{transp}, it is clear that the general
current density $j(Q)$ for arbitrary $Q$ (see Eq.~\eqref{curr}) is 
of order $O(1)$. In the following sections, we compute the transport of two 
important quantities.

\subsection{Particle current}
\begin{figure}
\begin{center}
\includegraphics{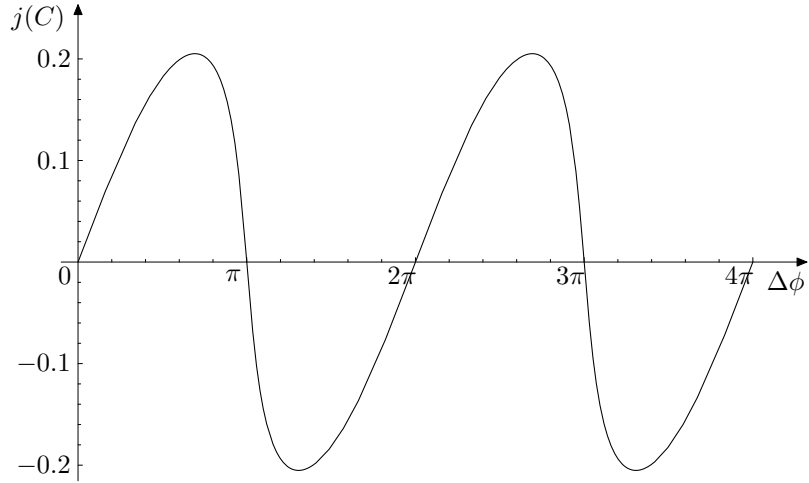}
\caption{Dependence of the Josephson current $j(C)$ 
on the phase difference $\Delta\phi$}
\label{fig3}
\end{center}
\end{figure}

Of course, the most interesting transport phenomenon 
in this model is the Cooper pair current through the junction. 
Is their really a current and what are its properties?
The relative particle current is defined as the transport of the 
relative number operator $C_N$
\begin{equation*}
C_N = \sum_{k_1,l_1=0}^{N-1} \si^+(k_1,l_1)\si^-(k_1,l_1) - 
\sum_{k_2,l_2=0}^{N-1}\sii^+(k_2,l_2)\sii^-(k_2,l_2)
\end{equation*} 
Calculating $j(C)$ using equation \eqref{transp}, we find
\begin{equation}\label{josephson}
j(C) =  2(\li\lti\sin(\phi_I-\tilde{\phi}_{I})-
\lii\ltii\sin(\phi_{II}-\tilde{\phi}_{II}) )
\end{equation}
The order of magnitude of the total flux of Cooper pairs is proportional to the 
size of the contact surface.  
It depends explicitly on the sines of the phase differences between the bulk and
the contact surface, \ie $\phi_i -\tilde{\phi}_i$ in both superconductors, $i =
I,II$, 
and on the order parameters, see \eqref{op1}--\eqref{op3}.
However, the only free parameters are the bulk phases
$\phi_I$ and $\phi_{II}$, the other parameters are determined by the 
selfconsistency equations \eqref{op2}--\eqref{op3}.
Hence, the interesting point here is the dependence of the Cooper pair current
on the phase difference between the two (bulk) superconductors  
$\Delta \phi = \phi_{II} -\phi_I$.
Let us illustrate this with an example. In Figure \ref{fig3}, we plotted the 
Josephson current $j(C)$
\eqref{josephson} as a function of the phase difference 
$\Delta \phi = \phi_{II} -\phi_I$.  
For this figure, we took $\beta_I = \beta_{II} = 4$, $\ei = \eii = 1/4$ and $\gamma =
1/2$.
If the phase difference is zero
$\Delta \phi = 0$, there is no Cooper pair current in the 
system, and if the sign of the phase difference is inverted, also the direction
of the superconducting current is reversed.
Hence, this supercurrent $j(C)$ \eqref{josephson} has the basic properties of the
superconducting tunnel current predicted by Josephson \cite{josephson:1962}. 
In the textbook derivations of the Josephson current, the current is proportional
to the sine of the phase difference $\Delta\phi$. In this model, however, the 
relation with the phase difference is more complex \eqref{josephson}, but as can
be seen in the example (Figure~\ref{fig3}), it has a similar sinusoidal 
dependence on the phase difference of the order parameters in both 
superconductors.

\subsection{Heat fluxes and entropy production}

Here, we want to make
a link with the current research about entropy production in quantum
statistical models \cite{ruelle:2001,jaksic:2002,ruelle:2002}. We investigate 
the entropy production in our model and compare different notions.

In quantum systems when the time evolution is given by a unitary
conservative dynamics, it is known that the von Neumann entropy does not 
change. This unitary time evolution however, induces correlations between 
different parts of our large system. It is natural to think of an observer who 
can only observe part of the large system. When after some time the correlations
of this part with other parts in the large system are forgotten, the entropy 
of this subsystem has changed, or entropy has been produced. 
In this way, entropy production can be understood as the change of
entropy when partitioning the total system in different parts \cite{ruelle:2002}.

Applying this idea to our model and considering the entropy production induced
by the partitioning of the total system into the two superconductors, we find 
the following expression for the entropy production \cite{ruelle:2002}:
\begin{equation}\label{e1}
e_1 = \lim_{N\to\infty} - \ii\frac{1}{N} \tilde{\omega}\big(\big[H_N,\log \left .
\tilde{\omega}\right |_{I} + \log \left . \tilde{\omega}\right |_{II}\big]\big),
\end{equation}
where $\left . \tilde{\omega}\right |_{I}$ and 
$\left . \tilde{\omega}\right |_{II}$ are the density matrices of the state
$\tilde{\omega}$ restricted to resp.\ the first and the second superconductor.

However, due to the product character of the steady state and using cyclic
permutations under the trace, it is
straightforward to check that the expectation value of this commutator vanishes. 
Hence, there is no entropy production in this model, \ie $e_1 =0$.

The definition \eqref{e1} can be compared with the conventional notion of 
entropy production, \eg see \cite{ruelle:2001,jaksic:2002},
where the entropy production is defined as the sum of the heat fluxes in the 
system, \ie we calculate the current \eqref{curr} associated with the following 
observable
\begin{equation*}
Q_N = \beta_I H_{I,N} + \beta_{II} H_{II,N},
\end{equation*}
where $\beta_I$ and $\beta_{II}$ are the initial inverse temperatures of the 
equilibrium states of the two superconductors and $H_{I,N}$ and $H_{II,N}$ are 
the BCS Hamiltonians \eqref{BCSHAM}. They represent the internal energy in the
superconductors. 
This leads to the following current \eqref{curr}:
\begin{eqnarray}
 \nonumber\lo{
e_2=j(Q) =} \lim_{N\to \infty} \frac{2}{N}\beta_I(\tilde{\omega}(\si^z(k_1,0))+2\ei)\li\lti\sin(\phi_I-\tilde{\phi}_I)
\\ \label{e2}
+ \frac{2}{N}\beta_{II}(\tilde{\omega}(\sii^z(k_2,0))+2\eii)\lii\ltii
\sin(\phi_{II}-\tilde{\phi}_{II})
\\ \nonumber
\lo{=}0.
\end{eqnarray}
Hence, the first definition \eqref{e1} of entropy production and the
conventional definition \eqref{curr} in terms of heat fluxes both lead to 
zero entropy production. 
However, for the last case, we see that the heat fluctuations are of 
order $O(N^{-1})$ around zero. Also, there is no definite sign for the entropy 
production fluctuations \eqref{e2} in this state as the sign depends on the 
phase differences, \ie just as in the case of the Josephson supercurrent 
\eqref{josephson}, but these heat currents are of a lower order of magnitude.  
This way, we may conclude that there are no heat fluxes in this system, 
there are only small fluctuations around zero. 

Finally, remark that in our setup, 
we took twodimensional superconductors and a onedimensional
contact surface. Of course, 
it might be interesting to look at other dimensions for the superconductors and
the contact surface.
We expect that our methods and results remain valid in those situations. 
The onedimensional case, however, is basically different.
If the contact surface is finite in the thermodynamic limit, the
nonequilibrium steady state does not factorise in the contact region.

\section*{Acknowledgements}
We wish to thank Christian Maes and Wojciech De Roeck for useful
discussions. 
J.~Lauwers gratefully acknowledges het Bijzonder Onderzoeksfonds K.U.Leuven 
for financial support.

\section*{References}

\bibliographystyle{unsrt}
\bibliography{biblio}

\end{document}